\documentclass[12pt]{article}
\usepackage{amsmath,latexsym,amssymb,epsfig,psfrag}
\newcommand{\be}{\begin{equation}}
\newcommand{\ee}{\end{equation}}
\input epsf
\usepackage{color} 

\newcommand{\I}{\mathrm i} 
\newcommand{\ba}{\begin{array}} 
\newcommand{\ea}{\end{array}}
\newcommand{\bqa}{\begin{eqnarray}}
\newcommand{\eqa}{\end{eqnarray}}
\newcommand{\nms}{\normalsize}
\newcommand{\abs}[1]{\left|#1\right|}
\newcommand{\tam}[1]{{\text {\Large $#1$}}}
\newcommand{\lb}{\left(}
\newcommand{\rb}{\right)}
\newcommand{\lsb}{\left[}
\newcommand{\rsb}{\right]}
\newcommand{\lcb}{\left\{}

\def\pd{\partial}

\def\di{\rm d}
\def\a{\alpha}
\def\b{\beta}

\def\g{\gamma}

\def\d{\delta}

\def\di{\mathrm{d}}

\def\m{\mu}
\def\n{\nu}

\def\t{\tau}
\def\th{\theta}

\def\th{\theta}
\def\l{\lambda}
\def\L{f}
\def\W{\Omega}

\def\sig{\sigma}
\def\Sig{\Sigma}
\def\e{\epsilon}

\def\ta{\mathrm {tan}}
\def\sn{\mathrm {sin}}
\def\cs{\mathrm {cos}}

\def\cth{\mathrm {coth}}

\def\id{1}
\def\nr{\frac{n}{R}}

\def\rfh{\mathfrak{e}_{f_h}}
\def\rfn{\mathfrak{e}_{f_\nu}}

\makeatletter   
\@addtoreset{equation}{section}   
\makeatother   
   
\def\simlt{\stackrel{<}{{}_\sim}}
\def\simgt{\stackrel{>}{{}_\sim}}  
\begin{document}

\setlength{\oddsidemargin}{0cm}
\setlength{\baselineskip}{7mm}

\begin{titlepage}  

\vskip 2cm
\begin{flushright}
{\bf UAB-FT-643}
\end{flushright}
\begin{center}  
\vspace{0.5cm} \Large {\sc Dirac Vs.~Majorana Neutrino Masses\\
\vspace{.5cm}  From a TeV Interval}
\vspace*{10mm}  
\normalsize  
  
{\bf 
D.~Diego~\footnote[1]{E-mail: diego@ifae.es}$^{,\,a}$
and
M.~Quir\'os~\footnote[2]{E-mail: quiros@ifae.es}$^{,\,a,\,b}$ 
}   

\smallskip   
\medskip   
\it{$^{a}$~Theoretical Physics Group, IFAE/UAB}\\ 
\it{E-08193 Bellaterra (Barcelona) Spain}

\smallskip     
\medskip
\it{$^b$~Instituci\'o Catalana de Recerca i Estudis Avan\c{c}ats (ICREA)}

\vskip0.6in \end{center}  
   
\centerline{\large\bf Abstract}  

\noindent
We investigate the nature (Dirac vs.~Majorana) and size of left-handed
neutrino masses in a supersymmetric five-dimensional model
compactified in the interval $[0,\pi R]$, where quarks and leptons are
localized on the boundaries while the gauge and Higgs sectors
propagate in the bulk of the fifth dimension. Supersymmetry is broken
by Scherk-Schwarz boundary conditions and electroweak breaking
proceeds through radiative corrections.  Right-handed neutrinos
propagate in the bulk and have a general five-dimensional mass $M$,
which localizes the zero modes towards one of the boundaries, and
arbitrary boundary terms.  We have found that for generic boundary
terms left-handed neutrinos have Majorana masses. However for specific
boundary configurations left-handed neutrinos are Dirac fermions as
the theory possesses a conserved global $U(1)$ symmetry which prevents
violation of lepton number. The size of neutrino masses depends on the
localization of the zero-modes of right-handed neutrinos and/or the
size of the five-dimensional neutrino Yukawa couplings. Left-handed
neutrinos in the sub-eV range require either $MR\sim 10$ or Yukawa
couplings $\sim 10^{-3}R$, which make the five-dimensional theory
perturbative up to its natural cutoff.
     
\vspace*{2mm}   
  
\end{titlepage}  
\section{\sc Introduction}
\label{introduction}

Within the Standard Model (SM) framework the origin, nature and
lightness of neutrinos are still open questions. A commonly accepted
explanation for Majorana neutrinos is the so called seesaw
mechanism~\cite{Minkowski:1977sc,Mohapatra:1979ia}.  It requires the
existence of a sterile~\footnote{With respect to the SM gauge group.}
and massive ($M_R\gg M_Z$) right-handed (RH) spinor weakly coupled to
the SM left-handed (LH) neutrinos and the Higgs through a Yukawa term
yielding thus a light Dirac mass ($m_D \ll M_R$). The lowest mass
eigenvalue is then $m_\n\sim\frac{m_D^2}{M_R}$ and it is $\sim meV$ if
the Yukawa coupling is weak enough and/or the RH neutrino mass is
large enough. It was originally proposed in the context of
four-dimensional (4D) models and it predicts the existence of new
physics at the scale $M_R$. The drawback of the seesaw mechanism is
that it either requires an extremely large value of $M_R$ [out of
reach for the Large Hadron Collider (LHC)] or introducing a small
value of the 4D Yukawa coupling $h_\nu\sim 10^{-7}$. The alternative
possibility that neutrinos are Dirac fermions would require much
smaller 4D Yukawa couplings $h_\nu\sim 10^{-14}$. In any way the
nature of neutrinos (Dirac or Majorana), which is essential for many
experimental signatures, still remains as the big open question in
neutrino physics.

Some of the above problems may find a solution in the context of
theories with (compactified) extra dimensions. They provide a natural
scale, the inverse compactification radius of extra dimensions $1/R$,
and moreover they can ``naturally'' provide very small 4D Yukawa
couplings if some of the fields participating in the localized Yukawa
interactions are exponentially localized far away. This led to the
physical scenario where the RH neutrino belongs to a hypermultiplet
propagating in the bulk of the extra dimensions while Yukawa
interactions are localized at fixed points of the bulk, e.g.~in
orbifold compactifications.

In general in a theory defined in an extra dimensional scenario two
main points should be answered:
\begin{itemize}
\item
Why neutrinos are so light?
\item
What is the nature of neutrinos: Dirac vs.~Majorana fermions?
\end{itemize}
Although lots of different studies have been done so
far~\cite{Dienes:1998sb}-\cite{Gherghetta:2007au} in supersymmetric or
non-supersymmetric theories, and flat or warped space-time a
systematic analysis of the previous questions was not addressed in a
compactified theory with arbitrary boundary terms.

In this paper we have investigated in detail the lightness and nature
of neutrinos in a supersymmetric five-dimensional (5D) theory defined
on the interval $\mathcal M_4\times [0,\pi R]$ where gauge and Higgs
bosons propagate in the bulk of the fifth dimension and SM fields
belong to chiral superfields localized in one of the boundaries of the
5D space. We will also consider the compactification scale in the TeV
range, which can lead to production of KK
excitations~\cite{Antoniadis:1990ew,Antoniadis:1994yi} at the
LHC~\footnote{This possibility will of course depend on the detailed
localization properties of quarks and leptons.} and where
supersymmetry can be globally broken by the Scherk-Schwarz (SS)
mechanism~\cite{Scherk:1978ta}. In these theories the supersymmetric
spectrum has a very characteristic pattern~\cite{Pomarol:1998sd},
which can very easily be identified in experiments, and electroweak
symmetry breaking (EWSB) can proceed by radiative
corrections~\cite{Diego:2005mu,DGQ} bypassing some previously outlined
difficulties in Ref.~\cite{Barbieri:2003kn}.

The plan of the paper is as follows. In section~\ref{Theaction} the
most general supersymmetric action for RH neutrinos in the bulk is
presented as well as the boundary terms, and boundary conditions
arising from the variational principle are written. In
section~\ref{wave} the wave functions for the bosonic and fermionic
components of the RH neutrino hypermultiplet satisfying the equations
of motion are deduced and the number of degrees of freedom
identified. In section~\ref{effaction} the 4D theory obtained after
integrating the fifth dimension is worked out and the effective action
for the light neutrino is computed in the background of the Higgs
field. Our main physical results can be found in
section~\ref{phenomenological} where the different situations
concerning the nature and size of LH neutrino masses are discussed.
Finally section~\ref{conclusions} contains our conclusions.

 A short summary of the main results is appended now. Concerning the
nature of the LH neutrino mass we have found that everything depends
on the relative orientation of the boundary action (characterized by a
vector $\vec s$) with respect to the bulk action of the RH neutrino
characterized by a mass vector $\vec p$. In the generic situation
where both vectors $\vec s$ and $\vec p$ are arbitrarily oriented,
lepton number is violated and we find a Majorana mass for LH
neutrinos. In the particular case where $\vec s$ and $\vec p$ are
parallel (or anti-parallel) there is a conserved global $U(1)$
symmetry group which prevents lepton number breaking and produces
Dirac masses for neutrinos. As for the size of the LH neutrino masses,
in the case of Dirac masses they can be in the meV range if the
zero-mode of the RH neutrino is exponentially localized towards the
brane opposite to that where the SM fermions localize. In particular
if $M$ is the localizing mass of RH neutrinos the correct size for the
LH neutrino mass requires that $MR\sim 10$. In the generic case where
LH neutrino masses are Majorana, in order to get the correct size
without imposing anomalously small 5D (dimensional) Yukawa couplings
it is shown that quarks and leptons should be localized in opposite
boundaries while the bulk Higgs localizes towards the quark boundary.
In this case the Yukawa couplings should be somewhat small ($\sim
10^{-3} R$) although for particular values of $\vec s\cdot\vec p$
there is a cancellation in the neutrino mass matrix for large values
of $M$ and the correct spectrum of neutrinos can be again obtained for
$MR\sim 10$.

\section{\sc The action and the spectrum}
\label{Theaction}

The action for the sterile neutrino is that of a 5D $N = 1$
supersymmetric theory with boundary terms defined in the manifold
$\Sig = M_4 \times I$, where $M_4$ is the Minkowski space and $I$ is
the interval $\left[0, \pi R\right]$, $R$ being the compactification
radius that we assume to be $\simlt{\rm TeV}^{-1}$. Thus the right
handed neutrino belongs to a (SM singlet) hypermultiplet $(N,N_c)$
with $N_{(c)} = \phi_{(c)} + \sqrt{2} \theta \psi_{(c)} + \theta^2
F_{(c)}$. 

The explicit form of the action is
\be
S = S_{\rm bk} + S^{\rm m}_{\rm bk} +  S_{\rm bd} \label{action}\,,
\ee
where
\bqa
           S_{\rm bk}  & =  &  
\int_\Sig d^4 \theta\,\left[\bar N N + 
\bar N_c N_c\right] - \int_\Sig d^2 \theta\,N_c\pd_5 N + {\rm h.c.} \,, 
\label{bklagrangian} \\
          S^{\rm m}_{\rm bk} & = & 
\int_\Sig d^2\theta \left[a N_c\,N + 
\frac{b}{2} N^2 - \frac{b^*}{2} N^2_c\right] + {\rm h.c.}\,, 
\label{masslagrangian}\\
           S_{\rm bd} &=& \int_{\pd\Sig}d^2 \theta\,\left(\frac{\mu}{2}
           N^2 + \frac{\lambda}{2} N^2_c + \nu N N_c\right) + {\rm
           h.c.}\,, \label{bdlagrangian}
\eqa
and $\mu ,\, \lambda$ and $\nu$ are arbitrary (dimensionless) complex
numbers~\footnote{It is understood that boundary terms occur
independently at $y = 0$ and $y = \pi R$.}, while the bulk mass terms
only depend on two (mass) parameters ($a \in \mathbb R$, $b \in
\mathbb C$) in order to guarantee the $SU(2)_R$ invariance. In fact
the fermionic component of Eq.~(\ref{masslagrangian}) provides the
most general fermion mass Lagrangian invariant under the 5D Lorentz
transformations
\begin{equation}
a \psi_c \psi +\frac{b}{2}\psi \psi-\frac{b^*}{2}\psi_c\psi_c+h.c=
a\bar\Psi\Psi+\frac{b}{2}\bar\Psi\Psi^c+\frac{b^*}{2}\bar\Psi^c\Psi
\end{equation}
where $\Psi$ is the Dirac spinor
\be
\Psi=\left(
\begin{array}{c}
\psi_c\\
\bar\psi
\end{array}
\right)
\label{Diracfermion}
\ee
$\Psi^c=C_5\bar\Psi^T$ its charge conjugate and
$C_5=\gamma_2\gamma_0\gamma_5$ the 5D charge conjugation
operator~\footnote{Note the different definitions of the 5D ($C_5$)
and 4D ($C_4=-i\gamma_2\gamma_0$) charge conjugation operators which
gives rise to the minus sign which appears in
Eq.~(\ref{masslagrangian}). }.

By applying the variation principle on
(\ref{bklagrangian})-(\ref{bdlagrangian}) we find the boundary
conditions
\be \mu N + \nu N_c - N_c = 0\,,\qquad \lambda N_c + \nu N = 0
         \,,\label{BC} 
\ee 
which turn out to be a set of manifestly supersymmetric boundary
conditions.  One can easily check that the system of equations
(\ref{BC}) is overdetermined unless
   \begin{equation}\label{overdetermine}
           \mu\,\lambda - (\nu -1) \nu = 0 \,,
  \end{equation}
and that it is invariant under the redefinitions 
\be
   N_c\leftrightarrow N, \, \l\leftrightarrow\m,\,\n\leftrightarrow 1-\n\,.
\label{relaciones}
\ee
In the special case $\n = 0$ the boundary conditions read
\be
   \lcb\ba{l}\l = 0,\,\, \m\,N - N_c =0\\\hspace{1cm}{\rm or}\\
                    \m =0,\,\, N_c = 0\ea\right.\,.
\ee 
while the case $\nu=1$ is obtained from the previous one by means of
the relations (\ref{relaciones}).  In the general case where $\n\neq
0,1$ Eq.~(\ref{BC}) reduces to
\be
\frac{\l}{\n} N_c + N = 0\,,
\ee
which means that the complex parameters $\left(\n,\m,\l\right)$ are
highly redundant since only the complex number $z = \l/\n$ does
matter. Notice that by letting $z$ to take any complex value we cover
the whole set of boundary conditions (including $\n=0$, which
corresponds to $z\to\infty$). Actually one can easily show that the
whole complex plane is covered consistently with (\ref{overdetermine})
by the restriction
$$\n \in \mathbb R\,,\qquad \m = - \l^*\,.$$         
As a matter of fact a possible parametrization is given by
\be \n = \frac{1}{2}\left(1 + s_3\right)\,,\qquad\m = -\l^* =
   \frac{1}{2}s_-\,,\label{param} 
\ee
with $s_- = s_1 - \I s_2$ where $s_i$ are three real parameters such
that $\vec s^{\;2} = 1$. 

The boundary conditions now read as
\be \frac{1}{2}\left(1- \vec s\cdot\vec\sig\right)
\left(\ba{c}N_c\\N\ea\right) = 0\,,\label{susybc}
\ee
where $\vec\sig$ are the Pauli matrices. In components
Eq.~(\ref{susybc}) reads as
\bqa
  \frac{1}{2}\left(\id - \vec s\cdot\vec\sig\right) \left(\ba{c}\phi_c\\
\phi\ea\right)& = & 0\nonumber\,,\\
  \frac{1}{2}\left(\id + \vec s\cdot\vec\sig\right)
\left[\pd_y + \vec p\cdot\vec s \, M\,\right] \left(\ba{c}\phi_c\\
\phi\ea\right)& = & 0\,, \label{Sbc} \\
  \frac{1}{2}\left(\id - \vec s\cdot\vec\sig\right)\left(\ba{c}\psi_c\\
\psi\ea\right) & = & 0\,,\label{Fbd}
\eqa 
respectively, where we have defined the shorthands
\be
\vec p = \frac{1}{\sqrt{a^2 + \abs{b}^2}}(b_R,- b_I,a)\,,
\qquad M = \sqrt{a^2 + \abs{b}^2}\,.\label{shand1}
\ee

The spectrum allowed by these boundary conditions can be read off from
Refs.~\cite{Diego:2005mu,DGQ}, and it is provided be the zeroes of the
function
\be
    \sn^2(\pi\t) - (c_0-c_\pi)\frac{M}{\W}\ta(\pi\W\,R) - 
\left[\cs^2(\pi\t) + c_0c_\pi\frac{M^2}{\W^2}\right]\ta^2(\pi\W\,R)\,,
\label{spectrum}
\ee 
with $\cs(2\pi\t) = \vec s_0\cdot\vec s_\pi$, $c_f = \vec p\cdot\vec
s_f$ and $\W^2 = m^2 - M^2$, $m$ being the physical mass eigenvalue.
In particular for $c_0 = c_\pi\equiv c$ ($\t = 0$) the spectrum
predicted by (\ref{spectrum}) is
 \be
     m^2 = s^2 M^2\,,\qquad m_n = \frac{1}{R}\,\sqrt{M^2 R^2 + n^2}\,,\,\,
n = 1,2,3, \cdots \label{solution}
 \ee
with $s^2 = 1-c^2$. 

\section{\sc Wave functions}
\label{wave}

In this section we will explicitly write the solutions to the
equations of motion and the boundary conditions for the model
presented in the previous section.  Let us first re-express the action
in a more appropriate way. As one can see from (\ref{masslagrangian})
the most general mass term involves both Dirac $(a)$ and Majorana
$(b,b^*)$ masses and there is a family of continuous transformations
parameterizing the possible mass configurations. These are $SU(2)$
rotations acting in the space of chiral supermultiplets
$(N_c,N)^T$~\footnote{Notice that this symmetry makes sense since
superfields $N$ and $N_c$ are uncharged.}. Although those
transformations do not leave the action invariant they are symmetries
of the spectrum equation (\ref{spectrum})~\footnote{Actually the
spectrum equation does only depend on
$SU(2)$-invariants. Geometrically speaking these (global) unitary
transformations change the basis where the bulk and boundary matrices
are expressed but the relations between them remain unaltered.}.
According to (\ref{shand1}) the action (\ref{action}) can be written
as
\bqa
\mathcal S & = & \int_{\Sig}\,\,\left.\bar{\mathcal N}\,{\mathcal N}
\right|_{\bar\th^2\th^2} - \frac{1}{2}M\left.
{\mathcal N}^T\tam{\e}\,\vec p\cdot\vec\sig\,{\mathcal N}
\right|_{\th^2} 
- \frac{1}{2}\left.{\mathcal N}^T{\text {\Large $\e$}}\,
{\mathcal N}'\right|_{\th^2} + {\rm h.c.} \nonumber \\
& + & \frac{1}{4}\int_{\pd\Sig}\,\,\left.{\mathcal N}^T\tam{\e}
\left(\id - \vec s\cdot\vec\sig\right)\,{\mathcal N}\right|_{\th^2} + 
{\rm h.c.} \,,\label{redef}
\eqa       
where 
\be\mathcal N = \left(\ba{c} N_c\\ N\ea\right)\,,\ee
and $\tam{\e}$ is the totally antisymmetric 2-tensor defined as
\be
\tam{\e} \equiv \lb\ba{cc}0&1\\-1&0\ea\rb = \I\sig_2\,.
\ee
For simplicity we will be interested in a pure Dirac bulk mass term,
i.e. $\vec p = (0,0,1)$. The set of unitary rotations bringing $\vec
p$ to $(0,0,1)$ is given by $U(\vec p)$
where
\begin{equation} 
U(\vec p) = \frac{1}{2\sqrt{1 - s_{\vec p}}}
\left(\ba{cc} e^{\I\d_{\vec p}}& 0 \\ 0 & e^{-\I\d_{\vec p}}\ea\right)
\left(\ba{cc} 1-s_{\vec p}+c_{\vec p} & 1-c_{\vec p}-s_{\vec p} \\ 
-1+c_{\vec p}+s_{\vec p} & 1+c_{\vec p}-s_{\vec p}\ea\right)\,,
\end{equation}
with $c_{\vec p} = p_3$, $s_{\vec p} = \sqrt{1-\lb p_3\rb^2}$ and
$$\tam{e}^{\I \d_{\vec p}} =(2 s_{\vec p})^{-1/2}\left(\sqrt{p_1+s_{\vec
p}}-\I \frac{p_2}{|p_2|}\sqrt{-p_1+s_{\vec p}}\right) $$
Notice that $U(p_1,p_2,0)$ is the limit of $U(\vec p)$ when $p_3 \to
0$, i.e.
\begin{equation}
U(p_1,p_2,0) = \frac{1}{\sqrt 2}\left(\ba{cc} e^{\I\d_{p_3 =0}}& 0 \\ 
0 & e^{-\I\d_{p_3 =0}}\ea\right)
\left(\ba{cc} 1 & -1 \\ 1 & 1\ea\right)\,.
\end{equation}
Under these rotations the kinetic term remains invariant while the
boundary matrices, like the bulk mass, transform covariantly.

In the basis where the bulk mass is Dirac-like we have
$\vec s_0=\vec s_\pi =(- s,0,c)$. The equations of
motion now read
\bqa
  \left.\ba{c}\left(\pd_5^2 - \Box - M^2\right)\,\phi = 0 \\
              \left(\pd_5^2 - \Box - M^2\right)\,\phi_c = 0 \ea\right\} & 
\textrm{for bosons,}\label{bosons}\\
    \nonumber \\
  \left.\ba{c}\I\,\bar\sig^\m\pd_\m\psi - \pd_5\bar\psi_c - M\bar\psi_c = 0 \\
                \I\,\bar\sig^\m\pd_\m\psi_c + \pd_5\bar\psi - M\bar\psi = 0 
\ea\right\} & \textrm{for fermions.}\label{fermions}
\eqa
\subsection{\sc Bosonic solutions} 

For a given mass $m$ the bosonic fields satisfy the Klein-Gordon
equation $\Box\,\phi_{(c)} = -m^2\phi_{(c)}$. Therefore the general
solution to the 5D equations of motion is given by 
\be \Phi (x,y)
= A(x)\,\cs(\W y) + B(x)\,\sn (\W y)\,,\label{bos} 
\ee
where $\Phi = \left(\ba{c}\phi_c\\ \phi\ea\right)$ and $\W$ has been
defined in (\ref{spectrum}). The boundary conditions (\ref{Sbc}) at $y
= 0$ impose the restrictions
\bqa
   A &=& \left(\ba{c} 1+c-s\\ 1-c-s\ea\right)\,a(x)\,,\nonumber\\
   B &=&\frac{c M}{\W}\left(\ba{c} 1+c-s\\ 
1-c-s\ea\right)\,a(x) + \left(\ba{c}-1+c+s\\ 
1+c-s\ea\right)\,b(x)\,, \nonumber
\eqa
where $a(x)$ and $b(x)$ are independent complex functions verifying
the above Klein-Gordon equation. Finally the boundary conditions at $y
= \pi R$ impose
\begin{align}
 & b(x)\, \sn(\W \pi R) = 0\,,\\
 & a(x)\left[\W + \frac{c^2M^2}{\W}\right] \sn(\W\pi R) = 0\,,
\end{align}
with two possible solutions
\begin{enumerate}
\item
\label{boszeromode} 
{\nms $\sn (\W\pi R)\neq 0$} and hence $b(x)=0$ and {\nms $\W^2= - c^2
M^2$}, whose eigenfunction is
\begin{align}
         \Phi^0  =& \left(\ba{c} 1+c-s\\ 
1-c-s\ea\right)\,\left[\cs(\W R y) - 
\frac{c\,M}{\W} \sn(\W R y)\right]\varphi(x)  \nonumber\\
         = &\left(\ba{c} 1+c-s\\ 
1-c-s\ea\right)\,\tam{e}^{-M cR y}\varphi(x)\,,
       \end{align}
\item
\label{ferzeromode} 
{\nms $\W R = n \in \mathbb Z_+$}, with eigenfunction
\bqa
  \Phi^{n} & = &  \left(\ba{c} 1+c-s\\ 
1-c-s\ea\right)\,f^n(y)\,\varphi^n_1(x) \nonumber\\
            &+& \left(\ba{c}-1+c+s\\ 1+c-s\ea\right)\,g^n(y)\,\varphi^n_2(x)
\eqa
with {\nms $f^n(y) = \cs(\nr y) - \frac{ M c R}{n}\sn(\nr y)$} and
{\nms $g^n(y) = \sn(\nr y)$}.
\end{enumerate}

 \subsection{\sc Fermionic solutions}

Eq.~(\ref{fermions}) can be re-casted into a single Dirac equation as
\be\label{diraceq}
  \left(\I\,\g^\m\pd_\m - \g^5\pd_5 - M\right)\,\Psi = 0\,,
\ee
with $\Psi$ the Dirac spinor defined in Eq.~(\ref{Diracfermion}), and
whose formal solution is given by
$$\Psi = \left[\cs(\sqrt{-\Box - M^2}\,y) + 
\g^5\frac{\I\,\g^\m\pd_\m - M}{\sqrt{-\Box - M^2}}\,
\sn(\sqrt{-\Box - M^2}\,y)\right]\,\Theta(x)\,,
$$
where $\Theta(x)$ is the initial value at $y = 0$ which fulfills the
4D Klein-Gordon equation $\Box\,\Theta + m^2\Theta = 0$. Thus the
eigenfunction corresponding to the $m$-th mode reads
\be
   \Psi = \left[\cs(\W y) + \g^5\frac{\I\,\g^\m\pd_\m - M}{\W}\,
\sn(\W y)\right]\,\Theta(x)\,.\label{formal}
\ee
Its initial value satisfies the boundary condition (\ref{Fbd}) at $y =
0$, that is:
\be 
\Theta = \left(\ba{c}(1+c-s)\chi\\(1-c-s)\bar\chi\ea\right)\,,\label{initial}
\ee
with $\chi$ an arbitrary 4D Weyl spinor. Since we are dealing with a
4D Dirac spinor, if $\xi$ is the Dirac partner of
$\chi$~\footnote{This is in straight analogy to the orbifold case
where we have a Dirac spinor in the bulk although parity assignment
projects out one of the components at the fixed points such that there
we have a single Weyl (Majorana) spinor.} satisfying the equations of
motion
\be  
\I\,\sig^\m\pd_\m\bar\xi = m \chi\,,\qquad \I\,\bar\sig^\m\pd_\m\chi = 
m \bar\xi\,,\label{dirac}  
\ee
by plugging now (\ref{dirac}) and (\ref{initial}) in (\ref{formal}) we
obtain the fermionic wave eigenfunction
\bqa 
   \Psi & = & \left(\ba{c}(1+c-s)f^\Omega_-(y)\,\chi\\ (1-c-s)f^\Omega_+(y)\,
\bar\chi\ea\right) \nonumber\\
\nonumber\\ 
& +&  m\,\left(\ba{c} (1-c-s)\,\xi\\ 
-(1+c-s)\,\bar\xi\ea\right)\frac{\sn(\W y)}{\W}\,,\label{presolution}
\eqa
where
\be f_{\pm}^\Omega(y)=\cs(\W y) \pm \frac{M}{\W}\sn(\W y)\,.\ee

Finally the boundary condition at $y = \pi R$ imposes the further
condition
\be
  \frac{\sn(\W\pi R)}{\W}\left(\ba{cc}1-c&s\\ 
s&1+c\ea\right)\cdot\left(\ba{cc} - M (1+c -s)&m (1-c -s)\\ 
M (1-c-s)&- m (1 + c-s)\ea\right)\cdot\left(\ba{c}\chi\\ 
\xi\ea\right)=0 \,,\nonumber
\ee
which is equivalent to
\be
  \frac{\sn(\W\pi R)}{\W}\left(\ba{cc}s\, M&m\\  
s\, M & m\ea\right)\cdot\left(\ba{c}\chi\\ \xi\ea\right) = 0\,.\label{pi}
\ee
Eq.~(\ref{pi}) again has two possible solutions:
\begin{enumerate}
\item 
$\W R \notin \mathbb Z$ with the solution 
\be m \xi + s \,M \chi = 0\,.\label{zero}
\ee
The condition (\ref{dirac}) implies now $\xi = \chi\equiv \eta$ which
corresponds to a Majorana spinor and
\be
       m = - s\, M\,.
\ee

 \item
\label{n} 
$\W R = n$ with $n \in \mathbb Z_+$. In this case we have two
independent spinorial degrees of freedom, $\xi^n$ and $\chi^n$,
degenerated in mass.
\end{enumerate}

Thus we re-encounter the spectrum (\ref{solution}) and the
corresponding wave functions turn out to be
\bqa
   \Psi^n & = & \left(\ba{c}(1+c-s)f^n_-(y)\,\chi^n\\ 
(1-c-s)f^n_+(y)\,\bar\chi^n\ea\right) \nonumber\\
\nonumber\\ 
& +&  \sqrt{n^2 + M^2 R^2}\,\left(\ba{c} (1-c-s)\,\xi^n\\ 
-(1+c-s)\,\bar\xi^n\ea\right)\frac{\sn(\nr y)}{n}\,,\label{n2}\\
\nonumber\\
\nonumber\\
  \Psi^0 & = & \left(\ba{c}(1+c-s)\eta\\
(1-c-s)\bar\eta\ea\right)\,e^{-c\,M y}\,. \label{0}
\eqa

Some comments about these solutions are now in order. On the one hand
notice that they are not, in general, factorizable as $f(y)g(x)$ which
makes it explicit that the orbifold-like {\it ansatz} is not
suitable~\footnote{In fact only in the particular case where $s=0$,
$c=\pm 1$ the solutions (\ref{n2}) break up into two orthogonal and
factorizable functions. Furthermore in the $M\to 0$ limit both
functions have a definite parity since the basis of mass
eigenfunctions is in that case $\{\cos (ny/R),\sin (ny/R)\}$. }.  On
the other hand we want to remark that the presence of $\xi$ does not
appear as a unitarity problem. From (\ref{dirac}) it can be expressed
in terms of $\bar\sig^\m\pd_\m\chi $ which is in agreement with the
uniqueness of the solution to (\ref{diraceq}). In fact $\xi$ and
$\chi$ can be thought of as off-shell independent degrees of
freedom. Finally from (\ref{formal}) and (\ref{bosons}) one
immediately checks that given a value of $\W$ the solutions
corresponding to $\pm\W$ are the same since there is no associated
degeneracy.  The sign of the root is a matter of convention since it
can be absorbed by the redefinition $y \to \pi R - y$, which shows
that both signs do correspond to the same eigenstate.

From now on we will concentrate on the fermionic sector describing RH
neutrinos. In particular we will write down the effective 4D action
for the hyperfermions $(\psi,\psi_c)$ coupled to the SM sector which
is localized at one of the boundaries, i.e.~$y = y_f$
($f=0,\,\pi$). The Yukawa couplings will induce a Dirac mass
connecting the left- and the right-handed neutrinos and we will find
the eigenvalues (Dirac or Majorana) of the effective mass matrix in an
extra dimensional generalization of the 4D see-saw
mechanism~\footnote{See Ref.~\cite{Dienes:1998sb} for a previous
non-supersymmetric analysis.}. In addition we will discuss in detail
the possibility of getting an ultra light mass eigenvalue with natural
values of the 5D Yukawa couplings.

\section{\sc Effective action for neutrinos}
\label{effaction}

In this section we will develop the 4D effective action for RH
neutrinos coupled to the SM matter localized at the brane
$y=y_{f_0}$. We will obtain the lowest mass eigenvalue in the neutrino
sector by solving the characteristic polynomial of the infinite
dimensional effective mass matrix involving the LH neutrino and the
zero and non-zero modes of RH neutrinos in the background Higgs
field. As we will show this can alternatively be done through the
effective mass matrix for the LH neutrino and the zero mode of the RH
neutrino resulting after integrating out the higher KK modes of RH
neutrinos. For the Higgs field which propagates in the bulk and whose
zero mode is exponentially localized towards one of the boundaries we
will consider the action computed in Ref~\cite{DGQ}.

The 5D action under study is~\footnote{The Yukawa interaction terms
are strictly localized on the boundaries since they are not $SU(2)_R$
invariant.}
\bqa
\mathcal S_{\rm eff} & = & 
\frac{1}{2} \int_{\Sig} \I \bar\psi_c\bar\sig^\m\pd_\m\psi_c  +  
\I \psi_c \sig^\m\pd_\m\bar\psi_c+ \I \bar\psi\bar\sig^\m\pd_\m\psi  +  
\I \psi \sig^\m\pd_\m\bar\psi\nonumber\\
  &-&\frac{1}{2}\int_{\Sig} \psi_c\left(-\pd_5\psi + M \psi\right) + 
\psi\left(\pd_5\psi_c + M \psi_c\right) + {\rm h.c.}\nonumber\\ 
& - & \frac{1}{4}\int_{\pd\Sig} \psi\left[s_- \psi + 
(s_3-1) \psi_c \right]+ \psi_c\left[(1+s_3) \psi - s_+ 
\psi_c \right] + {\rm h.c.} \nonumber\\
 &+& \int_{y =y_{f_0}} Y_\n\, 
\psi_c \n_L H_c + {\rm h.c.} \label{seesawaction}
\eqa       
where $\n_L$ denotes the LH neutrino, $Y_\n$ is the 5D Yukawa coupling
(with mass dimension -1), $y_{f_0}$ stands for the boundary where
$\nu_L$ is localized, i.e.~$f_0=0$ or $f_0=\pi$ and $H_c$ is the
lowest mode of the Higgs which according to Ref.~\cite{DGQ} is given
by $H_c \simeq \sqrt{2 M_H} \,e^{- M_H y}\,h(x)$~\footnote{The theory
of the Higgs hyperscalar ($H,\, H_c$) was worked out in detail in
Ref.~\cite{DGQ}. The bulk mass term in the Higgs sector $M_H$ is
further restricted by the $SU(2)\otimes U(1)$ invariance of the SM and
there is a similar parameter to $c_{0,\pi}$ defined in
section~\ref{Theaction}, $c_H$, which is restricted by the condition
of the Higgsino mass $\mu$ to be $s_H=\mu/ M_H$ so that $|c_H|\simeq
1$.}.  We now expand $\psi$ and $\psi_c$ as
\begin{align}
  \psi_c =& (1+c-s)e^{-M c y} \,\eta(x) \nonumber\\
+& \sum_{n\geq 1} \left[(1+c-s)f^n_-(y)\,\chi^n(x) + 
\frac{m_n}{n}(1-c-s)\,\sn(n\,y/R)\,\xi^n(x)\right]\ ,
\nonumber\\
\nonumber\\
\psi =& (1-c-s)e^{-M c y}\, \eta(x) \nonumber\\
 +& \sum_{n\geq 1} \left[(1-c-s)f^n_+(y)\,\chi^n(x) - 
\frac{m_n}{n}(1+c-s)\,\sn(n\,y/R)\,\xi^n(x)\right]\ ,\nonumber
\end{align}
and whose components satisfy the (free) equations of motion
\begin{align}
    &\I\bar\sig^\m\pd_\m\eta = -s\,M\,\bar\eta\,,\\
    &\I\bar\sig^\m\pd_\m\left(\ba{c}\chi^n\\ \xi^n\ea\right) = 
\frac{1}{R} m_n \left(\ba{cc} 0 & 1\\ 
1 & 0\ea\right)\cdot\left(\ba{c}\bar\chi^n\\ \bar\xi^n\ea\right)  \,,
\end{align}
with $m_n = \sqrt{M^2 R^2 + n^2}$.  

By integrating over the fifth coordinate one obtains the effective 4D
action
\begin{align}
 S_{\rm eff}= 
&\int\di^4x\,\lb k_0\rb^{-2}\left[\frac{\I}{2}\bar\eta\bar\sig^\m\pd_\m\eta + 
\frac{1}{2}s\,M\,\eta^2 \right]   \nonumber\\
 +& \sum_{n\geq 1}\int\di^4x\,
\left[\frac{\I}{2}\bar\L_nk_n\bar\sig^\m\pd_\m\L_n - 
\frac{1}{2 R} m_n\,\L^T_nk_n\sig_1\L_n \right]   \nonumber\\
  +& \int\di^4x\,Y_\n\,\sqrt{2M_H}\rfh\,(1+c-s)\,\eta\n_L h   \nonumber\\
  +& \sum_{n\geq 1}\int\di^4x\,Y_\n\,\sqrt{2M_H}\rfh (1+c-s)\,\chi_n\n_L\,h 
+ {\rm h.c.} \,,\label{effective}
\end{align}  
where $k_0$ is the real number
\be \lb k_0\rb^{-2} = \frac{2 (1-s)}{c M}\,\lb 1- 
\tam{e}^{-2 c M \pi R}\rb\,,\ee
$k_n$ the tower of matrices
\be 
k_n = 2\pi R\frac{\lb m_n\rb^2}{n^2} (1-s)\left[\id + 
s \frac{M R}{m_n}\,\sig_1\right]\,,
\ee
$\L_n$ stands for
$$\L_n = \left(\ba{c}\chi_n\\\xi_n\ea\right)\,,$$
and 
\begin{equation}
\rfh = \tam{e}^{- M_H |y_{f_h}-y_{f_0}|}.
\end{equation}
where $f_h=0$ or $f_h=\pi$ depending on the boundary the $H_c$ zero
mode is localized towards.  By redefining the modes as
 \begin{align} 
  \psi^n_\pm =  & \frac{1}{\sqrt 2\,k^{(n)}_\pm}\left( 
\chi^n \pm\xi^n\right)\,, \\
  \zeta =  & \frac{1}{k_0}\,\eta\,,
 \end{align}
 with 
 \be 
 (k^{(n)}_\pm)^{-2} = 2\pi R\frac{\lb m_n\rb^2}{n^2} (1-s)
\left[1 \pm s\frac{M R}{m_n}\right]\,,\label{kpm}
 \ee
the 4D effective action (\ref{effective}) can be rewritten as
\begin{align} 
S_{\rm eff}= &\int\di^4x\,\left[\frac{\I}{2}\bar\zeta\bar\sig^\m\pd_\m\zeta + 
\frac{1}{2}s\,M\,\zeta^2\right] + 
\sum_{n\geq 1}\int\di^4x\,\frac{\I}{2}\left(\bar\psi^n_-
\bar\sig^\m\pd_\m\psi^n_- + \bar\psi^n_+\bar\sig^\m\pd_\m\psi^n_+\right) 
\nonumber\\
  &- \sum_{n\geq 1}\int\di^4x\,\frac{1}{2 R} m_n\,
\left[\left(\psi^n_+\right)^2 - 
\left(\psi^n_-\right)^2\right] + \int\di^4x\,Y^{(0)}\zeta\n_L\,h \nonumber\\
&+\sum_{n\geq 1}\int\di^4x\,\left[Y^{(n)}_-\psi^n_-
\n_L\,h +Y^{(n)}_+\psi^n_+\n_L\,h\right] + {\rm h.c.} 
\label{effective2}\,,
\end{align}
where
\be Y^{(0)} = Y_\n\,(1+c-s)\, \sqrt{\frac{\abs{c M M_H}}{1-s}}\,\rfh\rfn
\label{Y0}
\ee 
and
\be Y^{(n)}_\pm =\frac{1}{\sqrt 2}Y_\n\,(1+c-s)M_H\,\rfh \,k^{(n)}_\pm\,,
\label{Ypm}
\ee
are the 4D effective Yukawa coupling constants. Here we have defined
$\rfn$ as
\be
   \rfn = \tam{e}^{-|c M (y_{f_\n}-y_{f_0})|}\,,
\ee
where $f_\n$ depends on where the lowest mode of the RH neutrino
localizes towards.  Once the Higgs gets its vacuum expectation value,
$\langle h \rangle= v$, the Yukawa couplings turn into Dirac mass
terms and we are thus left with an effective mass matrix connecting LH
and RH neutrinos given by
   \begin{align}
     \mathcal L_m =&\frac{1}{2}s\,M\,\zeta^2 + 
Y^{(0)} v\zeta\n_L- 
\sum_{n\geq 1}\frac{1}{2 R} m_n\,
\left[\left(\psi^n_+\right)^2 - \left(\psi^n_-\right)^2\right] \nonumber\\
+ &\sum_{n\geq 1} v 
\left\{Y^{(n)}_+\psi^n_+\n_L + Y^{(n)}_-\psi^n_-\n_L\right\}\,.
\label{effmass}
\end{align}
Notice that higher modes of RH neutrinos appear in (\ref{effective2})
as Majorana spinors albeit we started with Dirac fermions. However if
we redefine the fields for $n\neq 0$ as
\begin{align}
  \varphi^n_+ &= \psi^n_- + \psi^n_+ \,,\\
   \varphi^n_- &= \psi^n_- - \psi^n_+\,,
\end{align}
we recover the expected Dirac spinors due to the degeneracy in
mass of the higher modes.

Now we can find the eigenvalues of the infinite mass matrix
(\ref{effmass}) by computing its characteristic
polynomial~\cite{Dienes:1998sb}%
\be
P(\l)={\rm det}\left(\ba{cccccc}
-\l&m_D^{(0)}&\ldots &m_{D+}^{(n)}&m_{D-}^{(n)}&\ldots\\ 
   m_D^{(0)}& s\,M-\l\\
   \vdots &&\ddots\\
   m_{D+}^{(n)}&&&-m_n-\l\\
   m_{D-}^{(n)}&&&&m_n-\l\\
 \vdots&&&&&\ddots\ea\right)\,. \label{massmatrix}
\ee
where $m_D^{(0)} = vY^{(0)}$ and $m^{(n)}_{D\pm} = v Y_\pm^{(n)}$. The
determinant of (\ref{massmatrix}) yields
\begin{align}
&P(\l)=\\
 &\left[\left(\l- s\,M \right)\prod_{k \geq 1} 
\left(\l^2- m_k^2\right)\right]\left\{\l + \frac{\left(m_D^{(0)}\right)^2}
{s\,M-\l} + \sum_{n \geq 1} 
\left[-\frac{\left(m_{D+}^{(n)}\right)^2}{m_n +\l} + 
\frac{\left(m_{D-}^{(n)}\right)^2}{m_n - \l}\right]\right\}\nonumber
\label{caracterpol}
\end{align}

When $s\neq 0$ the smallest eigenvalue, $\l_L$, will not be that
canceling either $\l - s\,M$ or $\l^2-m^2_k$.  Therefore it should
verify the equation
\be
\l_L + \frac{\left(m_D^{(0)}\right)^2}{s\,M-\l_L} + 
\sum_{n \geq 1} \left[\frac{\left(m_{D-}^{(n)}\right)^2}{m_n -\l_L} - 
\frac{\left(m_{D+}^{(n)}\right)^2}{m_n + \l_L}\right] = 0\,.
\label{lambdalowest}
\ee     
Since $m_{D\pm}^{(n)}$ and $m_D^{(0)}$ are $\sim vY_\n/R$, and we are
assuming this parameter to be much smaller than $M$, we can expand the
solution for the dimensionless parameter $\lambda_L/M$ in powers of
$\b = \frac{v Y_\n} {M R}$ as
\be
\frac{\l_L}{M} = \sum_{\ell=1}^\infty\l_{2\ell}\b^{2\ell}\,.\label{betaserie}
\ee  
which makes sense whenever the lowest order is small.  Substituting
back in (\ref{lambdalowest}) we find that at lowest order $\l_L$ is
given by
\begin{align}
\frac{\l_L}{M} &+ \frac{\left(m_D^{(0)}\right)^2}{s\,M^2}
=-\sum_{n\geq 1}\frac{R}{M m_n}\left[\left(m^{(n)}_{D-} \right)^2 -
\left(m^{(n)}_{D+}\right)^2\right]\nonumber\\ =& -\frac{2(1+c)
s\,\lb\rfh\rb^2 R M_H v^2 Y_\n^2}{\pi} \sum_{n\geq 1}\frac{n^2}{(n^2 +
M^2R^2)(n^2 + c^2 M^2 R^2)}
\label{suma}
\end{align}
Finally the series in (\ref{suma}) can be computed analytically by
means of a Poisson re-summation giving
\begin{equation}
 \l_L =
  -  v^2 Y^2_\n\frac{1+c}{s}\,M_H \lb\rfh\rb^2 \left[2 c\lb\rfn\rb^2  + 
\cth(\pi M R) - c\,\cth(\pi c\,M R)\right]\,.
\label{ligero} 
\end{equation}
Notice that the series in (\ref{suma}) is finite because both
$m^{(n)}_{D\pm}$ converge to the same limit when $n\to\infty$. This
feature is just a consequence of the structure of the bulk mass matrix
in (\ref{masslagrangian}) which is consistent with 5D supersymmetry
and Lorentz invariance.

Alternatively this result can be obtained as the diagonalization of
the effective action induced after integrating out the higher KK-modes
$\psi^{n}_\pm$ in (\ref{effmass}) for momenta much smaller than their
masses (i.e.~neglecting their kinetic terms). From (\ref{effmass}) the
equations of motion for $\psi^n_\pm$ are given by
\be
\psi^n_\pm=\mp\frac{R \,m^{(n)}_{D\pm}} {m_n }
\, \n_L \ee
and substituting back in (\ref{effmass}) we find, in matrix form, the
following effective mass coupling for $\n_L$ and $\zeta\equiv\n_R$
\be
\frac{1}{2}\left(\n_L,\n_R\right)\cdot\left(\ba{cc}\m&m_D^{(0)}\\
m_D^{(0)}& s\, M\ea\right)\cdot\left(\ba{c}\n_L\\
\n_R\ea\right)\,,\label{2deffmass} 
\ee 
with 
\be \m =  \sum_{n\geq
1}\frac{R}{m_n} \,\left[\left(m^{(n)}_{D+}\right)^2
- \left(m^{(n)}_{D-}\right)^2\right] \,, 
\ee 
Since $\m$ and $m^{(0)}_D$ are much smaller than $M$ the light and
heavy eigenvalues $\lambda_{L,H}$ of the mass matrix in
(\ref{2deffmass}) are
\begin{equation}
\lambda_L\simeq\m - \frac{\left(m^{(0)}_D\right)^2}{s\,M}\,,\qquad
\lambda_H\simeq s\,M\,.  
\label{Maj}
\end{equation}
where the light eigenvalue $\lambda_L$ coincides with that found in
Eq.~(\ref{ligero}).

A particularly interesting case is found when $s = 0$, that is when
the boundary matrices are precisely aligned with the bulk mass
matrix. In that case the characteristic polynomial simplifies to
\begin{equation}
P_0(\lambda)=
\left[\prod_{k \geq 1} \left(\l^2- m_k^2\right)\right]
\left\{\l^2\left(1 + 2 \sum_{n\geq 1}\frac{ 
\left(m_{D}^{(n)}\right)^2}{m_n^2 - \l^2}\right) -  
\left(m_{D}^{(0)}\right)^2\right\}
\label{caracterpoldirac}
\end{equation}
where
\be
\left(m_{D}^{(n)}\right)^2 = 2\b^2 M^2 M_H R \frac{n^2}{\pi (n^2 + M^2 R^2)}\,,
\ee
Notice that (\ref{caracterpoldirac}) is an equation for $\l^2$ which
means that both $\pm\l$ are solutions and thus the set of eigenstates
of the whole mass matrix are exactly degenerate by pairs and therefore
they can be gathered to yield Dirac fermions.  Following the same
methods used above we find that the lowest eigenvalue is given by
 \be
 \l_{L\pm} =  \pm 2\,Y_\n v\,\rfh\sqrt{M_HM}  \,\tam{e}^{-\pi M y_{f_\n}}\,,
 \ee
The effective mass matrix for $\n_L,\n_R$ will be given in that case by
\be
\left(\ba{cc}0& m_D\\  m_D & 0\ea\right)\,.
\ee
where we have defined {\nms $m_D=2 Y_\n
v\,\rfh\sqrt{MM_H}\,\tam{e}^{-\pi M y_{f_\n}}$}.  

The degeneracy of the spectrum for $s = 0$ can be understood in terms
of a symmetry which takes place only within this case. As a matter of
fact $s=0$ means that the vectors $\vec p, \vec s_0, \vec s_\pi$ are
all aligned along the same direction and hence a $U(1)$ subgroup of
unitary rotations around this direction leaves the action
invariant. In terms of the fermion components these transformations
translate into
\be (\eta,\, \chi^n) \to
\tam{e}^{\I \a}\, (\eta,\, \chi^n) \,,\qquad \xi^{n}\to \tam{e}^{-\I
\a}\,\xi^{n}\,,\qquad \n_L \to \tam{e}^{-\I \a}\,\n_L 
\ee
where $\a$ is a real parameter. Notice that this symmetry forbids any
Majorana mass term~\footnote{This symmetry plays the role of the
lepton number symmetry of the SM.}, in particular for $\n_L$, and
hence $\m$ must vanish, as it is evident from Eq.~(\ref{effective}).

\section{\sc Discussion on neutrino masses}
\label{phenomenological}

In this section we will apply the previous results to discuss the
possibility of getting, within this kind of models, an (ultralight)
neutrino mass in the sub meV range. The first task will be to set the
range of dimensional Yukawa couplings which appear in the 5D action in
the leptonic sector
\be
\int_{\partial\Sigma} \left(Y_{\nu_\ell} H_c \n_L \psi_c+Y_{\ell}
H \ell_L e_R\right)
\ee
i.e.~$Y_{\nu_\ell}$, with mass dimension -1, and $Y_\ell$, with mass
dimension -1/2, where $\ell = \{\t,\m,e\}$. A naive estimate of wave
function renormalization corrections to the Yukawa couplings in the 5D
theory sets bounds as
\begin{equation}
y_{\nu_\ell}\equiv\frac{Y_{\nu_\ell}}{R}\simlt\frac{4\pi}{\Lambda R},\quad
y_{\ell}\equiv\frac{Y_{\ell}}{\sqrt{R}}\simlt\frac{4\pi}{\sqrt{\Lambda R}}
\end{equation}
so that taking $\Lambda R\sim 20$ we obtain $\mathcal O(1)$ upper
bounds on the dimensionless Yukawa couplings $y_{\nu_\ell,\,\ell}$.
We can now distinguish two different scenarios where neutrino masses
are either Dirac or Majorana:
\definecolor{blue}{rgb}{0, 0, 1}
\definecolor{red}{rgb}{1, 0, 0} %

\subsection{\sc Dirac mass}
We will assume here that all the SM matter is strictly localized on
the $y=0$ brane and the zero mode of the Higgs is localized towards it
as well, thus $\rfh = 1$, while the zero mode of RH neutrino is
exponentially localized towards $y=\pi R$, i.e. $y_{f_\n} = \pi R$ as
Fig.~1 shows.

\begin{center}
\begin{figure}[htb]
\setlength{\unitlength}{1.5mm} 
\begin{picture}(60, 50)(-40,-10)
\linethickness{0.5mm}
\put(0,0){\line(0,1){30}}
\put(-2,-5){{\footnotesize $y = 0$}}
\put(38,-5){{\footnotesize $y = \pi R$}}
\put(-8,36){{\footnotesize $Q_L, U_R, D_R $}}
\put(-5,32){{\footnotesize $L_L, E_R$}}
\put(40,0){\line(0,1){30}}
\linethickness{0.15mm}
\color{blue}
\qbezier(-1,25)(5,4)(38.6,2)
\put(3,20){{\footnotesize $H_c $}}
{\color{red}
\qbezier(-1,2)(35,4)(38.5,25)
\put(30,20){{\footnotesize $\n_R^{(0)}$}}}
\end{picture} 
{\footnotesize 
\caption{Bulk and brane matter distribution for a neutrino Dirac-like mass.}}
\label{diracneutrino}
\end{figure}
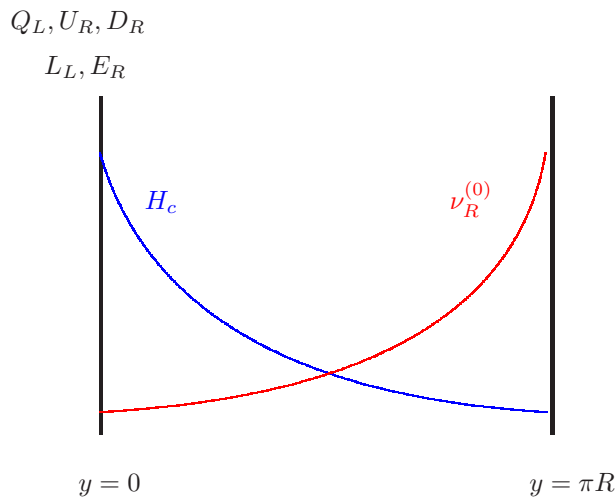
\end{center}

\begin{figure}[htb]
\begin{center}
\hspace*{-0.0cm}
\psfrag{MR}[][l]{$MR$}
\psfrag{mDR}[][b]{$\log_{10}{\displaystyle
\frac{m_{\n_\ell}^D}{\rm eV}}\hspace{1cm}$}
\includegraphics[width=86mm]{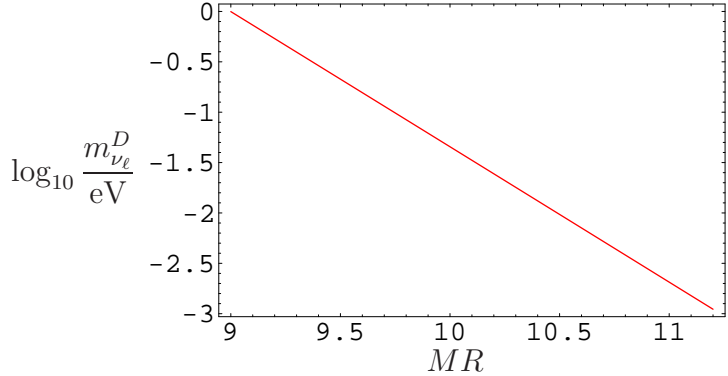}
\end{center}
\vspace*{-0.4cm}
\caption{\small $m_{\n_\ell}^D$ as a function of $M R$ for
$M_HR=1.6$ and $s=0$.}
\label{diracmass}
\end{figure}

We will consider for the moment the particular case where $s=0$,
i.e.~where vectors $\vec p$ and $\vec s_{0,\pi}$ are all aligned along
the same direction. As it was shown in the previous section we obtain
a Dirac mass connecting $\n_L$ and $\n_R$, which is of order
%
$m_{\n_\ell}^D
\sim 2v\,Y_{\n_\ell}\sqrt{M\,M_H}\,\e_R\,,$
%
with {\nms $\e_R = \tam{e}^{- M\pi R}$}.
In Fig.~\ref{diracmass} we show the Dirac mass as a function of $M R$
for $1/R= 5$ TeV, $y_{\n_\ell}\sim 1$ and $M_H R\sim
1.6$~\cite{DGQ}~\footnote{As it is shown in Ref.~\cite{DGQ} for values
of $M_H R$ near $M_H R\sim 1.6$, when supersymmetry is globally broken
by Scherk-Schwarz boundary conditions the spectrum of the Higgs
presents a tachyon at the tree level, which partially cancels the
positive one-loop radiative correction to the Higgs mass due to the
gauge coupling and allows EWSB to take place at the two-loop level
with a modest amount of fine tuning.}. We can see from
Fig.~\ref{diracmass} that $m_{\n_\ell}^D\simlt 1$ eV for $MR\simgt 9$
although $m_{\n_\ell}^D$ decreases exponentially when $MR$ increases
and thus $m_{\n_\ell}^D\simeq 1$ meV for $MR\simeq 11$.  In this
scenario there is no wave function suppression for the charged leptons
whose Yukawa couplings should therefore be given by $y_\ell\sim
m_\ell/v$.

If $s\neq 0$ in this scenario the lowest eigenvalue is a Majorana mass
given by
\be m^M_{\n_\ell} \sim Y_{\n_\ell}^2 v^2 M_H \lsb2c\,\tam{e}^{-2\pi
\abs{cM}R} + \cth(\pi M R) - c\,\cth(\pi c\,M R)\rsb\,, 
\ee 
which in general will be too large, unless there is a strong
suppression of the 5D Yukawa constants as {\nms $y_{\n_\ell}\sim
10^{-6}$}, which is similar to the electron Yukawa coupling in this
kind of models $y_e$.

Yet another possibility could be to localize the lowest mode of the RH
neutrino towards $y=0$, corresponding to $c M>0$. Considering now $M R
\gg 1$ the mass eigenvalue is proportional to $(c + {\rm sign}(M) +
2\,\tam{e}^{-2\pi \abs{M} R} - 2c\,\tam{e}^{-2\pi c\,M R})$.  Then by
choosing {\nms $c = - {\rm sign}(M)$} we could be left with an
exponentially suppressed Majorana mass. However this value of $c$ is
not consistent with the initial hypothesis $c M >0$. In fact the
smallness of the Majorana eigenmass is naturally achieved with a
different localization of the quark and lepton sector within the SM as
we will see in the next section.

\subsection{\sc Majorana mass}
The main obstruction to get a small Majorana mass eigenvalue out of
the effective mass matrix (\ref{2deffmass}) for the $s\neq 0$ case is
that the Yukawa couplings of the higher KK modes are not suppressed if
the SM matter is located on the same boundary where the Higgs
localizes towards. However by allowing the Standard Model matter to be
split into different branes~\footnote{Such a splitting may find its
justification within the context of intersecting branes in string
theory. See e.g.~\cite{Uranga:2005wn} and references therein.}, for
instance quarks localized in the same boundary (quark brane) where the
Higgs localizes towards, and leptons localized in the opposite
boundary (lepton brane) as Fig.~3 shows, then by means of the small
exponential wave function factor $\rfh$, the whole tower of effective
Yukawa couplings will be exponentially suppressed by the Higgs
localization and so $\m$ will be.
\begin{center}
\begin{figure}[htb]
\setlength{\unitlength}{1.5mm} 
\begin{picture}(60, 50)(-40,-10)
\linethickness{0.5mm}
\put(0,0){\line(0,1){30}}
\put(-2,-5){{\footnotesize $y = 0$}}
\put(38,-5){{\footnotesize $y = \pi R$}}
\put(-8,32){{\footnotesize $Q_L,U_R,D_R$}}
\put(35,32){{\footnotesize $L_L,E_R$}}
\put(40,0){\line(0,1){30}}
\linethickness{0.15mm}
\color{blue}
\qbezier(-1,25)(5,4)(38.6,2)
\put(3,20){{\footnotesize $H_c $}}
{\color{red}
\qbezier(-1,2)(35,4)(38.5,25)
\put(30,20){{\footnotesize $\n_R^{(0)}$}}}
\end{picture} 
{\footnotesize \caption{Bulk and brane matter distribution for a
neutrino Majorana-like mass. The $\n_R$ propagates in the bulk with
mass $M$.}}
\label{majorananeutrino}
\end{figure}
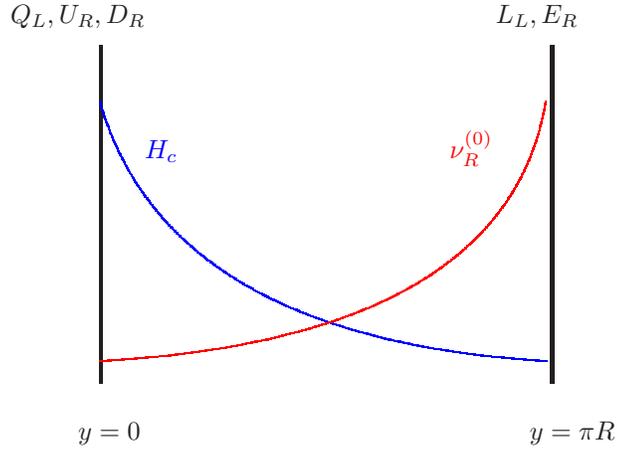
\end{center}
Now the effective mass matrix is analogous to the previous case except
for the global exponential suppression in the Dirac couplings, namely,
$\m \to \e^2_H\, \m$ with $\e_H = \tam{e}^{-\pi M_H R}$. In addition
we will assume the lowest mode of the RH neutrino to localize towards
the leptonic brane, i.e. $y_{f_\n} = 0$, corresponding thus to $c M <
0$. We then find that the lowest neutrino Majorana mass eigenvalue is
given by
\be
m_{\n_\ell}^M =  \e_H^2 v^2 Y_{\n_\ell}^2 M_H\frac{1+c}{s} 
\left[2c + \cth(\pi  M R) - c\,\cth(c\,\pi M R)\right]\,.
\label{majepsilonh}
\ee
Notice the almost independence on the RH neutrino bulk mass
$M$. Albeit its presence is absolutely necessary to provide the
existence of a lowest Majorana mass eigenvalue~\footnote{In case of
vanishing $M$ we would be left with a lowest Dirac eigenvalue or, at
most, with two almost degenerate Majorana eigenstates.} it is shielded
by the higher RH neutrino modes. In Fig.~\ref{masaMajorana} we plot
$m_{\n_\ell}^M$ as a function of $\log_{10} y_{\n_\ell}$ for fixed
values of $c$ and $MR$.
\begin{figure}[htb]
\begin{center}
\hspace*{-0.0cm}
\psfrag{z}[][l]{$-\log_{10}y_{\n_\ell}$}
\psfrag{mMR}[][b]{$\log_{10}{\displaystyle
\frac{m_{\n_\ell}^M}{\rm eV}}\hspace{1cm}$}
\includegraphics[width=96mm]{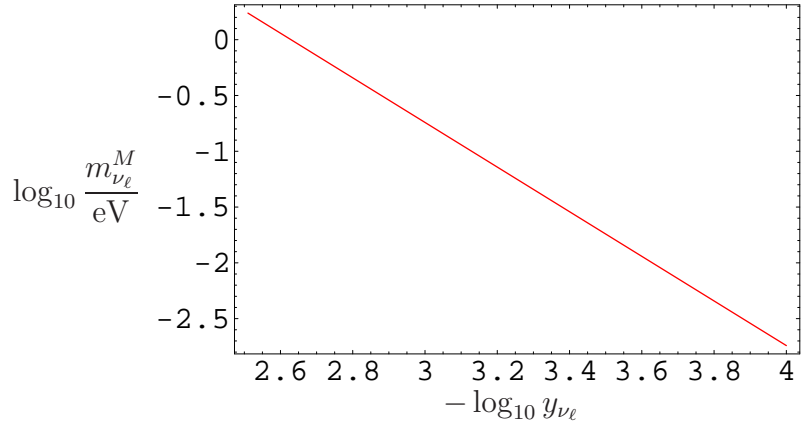}
\end{center}
\vspace*{-0.4cm}
\caption{\small Majorana neutrino mass, $m_{\n_\ell}^M$, as a function
of $-\log_{10}y_{\nu_\ell}$ for $c=-1/2$, $M_HR=1.6$ and $M R=5$.}
\label{masaMajorana}
\end{figure}

\noindent From Fig.~\ref{masaMajorana} we can see that generically
$m_{\n_\ell}^M\simlt 1$ eV implies $y_{\nu_\ell}\simlt 10^{-3}$.

The charged leptons, on the other hand, have masses 
\be m_\ell \sim v\,y_\ell \sqrt{M_HR}\,\e_H\,,
\ee
By fixing in this scenario the Higgs localizing mass to its previous
value $M_HR=1.6$ we can predict the correct value of the $\tau$
mass~\cite{Eidelman:2004wy} by means of a 5D Yukawa coupling
$y_\tau\simeq 1$ while $y_\ell\simeq m_\ell/m_\tau$ for the first two
generations $(\ell=e,\mu)$.

\begin{figure}[htb]
\begin{center}
\hspace*{-0.0cm}
\psfrag{w}[][l]{$|MR|$}
\psfrag{mMR}[][b]{$\log_{10}{\displaystyle
\frac{m_{\n_\ell}^M}{\rm eV}}\hspace{1cm}$}
\includegraphics[width=96mm]{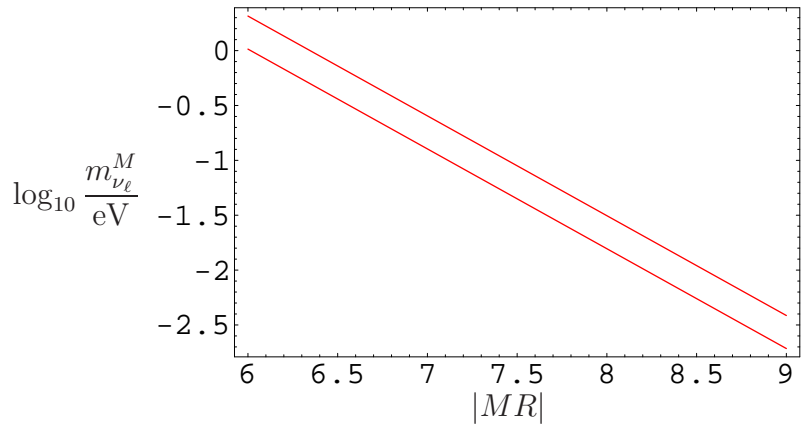}
\end{center}
\vspace*{-0.4cm}
\caption{\small Neutrino Majorana mass, $m_{\n_\ell}^M$, as a function
of $|MR|$ for $M_HR=1.6$ and $c=1/3$ (upper curve), $c=-1/3$ (lower
curve).}
\label{ultima}\vspace{1cm}
\end{figure}

An interesting particular case arises here. Given that $c M$ is
negative, in the limit when $\abs{c M R} \gg 1$ Eq.~(\ref{majepsilonh})
reads
\be 
m_{\n_\ell}^M\sim Y_{\n_\ell}^2v^2\e_H^2 M_H\frac{1+c}{s}\lsb 3c + {\rm
  sign}(M) + 2 {\rm sign}(M)\, \tam{e}^{-2\pi \abs{M}R} + 2 c\,
  \tam{e}^{-2\pi \abs{c M}R}\rsb\,.
\ee
Considering now the particular value {\nms $c = -\frac{1}{3} {\rm
sign}(M)$} we find a doubly suppressed Majorana mass eigenvalue. For
instance for the case $M>0$ and $c=-1/3$ one gets
\be 
m_{\n_\ell}^M \sim \frac{\sqrt{2}}{3} Y_{\n_\ell}^2v^2 \e_H^2
 \tam{e}^{-\frac{2}{3} \pi MR}\,.  
\ee
which has a doubly suppressed exponential behavior both from $M_HR$
and $MR$. In that case one can get tiny Majorana neutrino masses from
the localization of the zero mode of $\n_R$ for $\mathcal O(1)$ values
of the 5D Yukawa couplings $y_{\n_\ell}$. This is shown in
Fig.~\ref{ultima} where the Majorana mass $m_{\n_\ell}^M$ is plotted
versus $|MR|$ for $y_{\n_\ell}=1$ and $c=\pm 1/3$.

\section{\sc Conclusions}
  \label{conclusions}
In this paper we have investigated the nature and size of the LH
neutrino masses in a supersymmetric 5D model compactified in the space
$\mathcal M_4\times I$ where $I=[0,\pi R]$ is a finite interval of
length $\pi R$ and $R$ the compactification radius. Quarks and leptons
are localized on some of the boundaries and the gauge and Higgs sector
propagate in the bulk of the fifth dimension. In this model, as we
have found in previous works, supersymmetry can be globally broken by
a Scherk-Schwarz twist giving a mass to gauginos (and gravitinos)
which is transmitted by one-loop radiative corrections to squarks
localized on the boundaries, and provides a very characteristic
pattern for the supersymmetric spectrum which could be easily
identified experimentally at LHC whenever it is
found~\cite{Pomarol:1998sd}. Furthermore electroweak breaking proceeds
by radiative corrections~\cite{DGQ} providing at low energy an
MSSM-like 4D model.

We have found that the nature of the LH neutrino mass depends on the
relative orientation of the boundary terms, given by the vector $\vec
s$, with respect to the bulk mass term characterized by the vector
$M\vec p$. In the generic case of arbitrary orientations lepton number
is violated and neutrinos are Majorana fermions. In the particular
case where vectors $\vec s$ and $\vec p$ are aligned (or anti-aligned)
there is an extra global $U(1)$ symmetry which prevents lepton number
breaking and LH neutrinos get a Dirac mass. 

As for the size of the LH neutrino masses, in the case of Dirac
neutrinos, the smallness of neutrino masses is provided by the bulk
mass $M$ which should localize the zero mode of RH neutrinos towards
the opposite boundary to that where SM fermions are localized: sub-eV
masses are obtained for $MR\sim 10$. In the cases where neutrino
masses are Majorana, in order to avoid introducing too small neutrino
(dimensional) Yukawa couplings we have to localize quarks and leptons
on different boundaries. In that case sub-eV neutrino masses are
provided in general for Yukawa couplings $\sim 10^{-3} R$. However for
particular values of $\vec s\cdot \vec p$ there is a cancellation in
the neutrino mass matrix in the limit of large localizing masses and
the correct values $\sim$ meV simply require $MR\sim 10$.
 
Following the lines of our calculation it should be easy to describe
textures of RH neutrino masses describing the different patterns for
LH neutrino masses and mixings (see
e.g.~\cite{GonzalezGarcia:2002dz}). It should be enough to introduce
the corresponding three-by-three mass matrices in the 5D bulk and
boundaries and to carry on the parallel calculation. In the cases
where Dirac or Majorana neutrino masses are controlled by the
localizing masses of the RH neutrinos, since the former depend
exponentially on the latter a modest change in the corresponding RH
mass eigenvalues should be able to describe realistic neutrino
spectra. Of course since the RH neutrino masses are an input in our
theory, even if correct spectra do not require to fine-tune any
parameters, we should not call this a ``solution to the neutrino mass
problem'' until some more fundamental theory (e.g.~string theory)
would give us the correct values for the heavy masses. In fact this
was not the aim of our work but rather a classification of the
different solutions of the 5D theory providing realistic spectra for
neutrino masses and yielding hints for possible future discoveries at
LHC.

\subsection*{\sc Acknowledgments}
Work supported in part by the European Commission under the European
Union through the Marie Curie Research and Training Networks ``Quest
for Unification" (MRTN-CT-2004-503369) and ``UniverseNet"
(MRTN-CT-2006-035863); by the Spanish Consolider-Ingenio 2010
Programme CPAN (CSD2007-00042); and by CICYT, Spain, under contract
FPA 2005-02211.  We would also like to thank G.~von Gersdorff for
useful discussions and remarks and for initial collaboration in this
work, and to C.~Biggio for discussions.


\begin{thebibliography}{99} 


\bibitem{Minkowski:1977sc}
  P.~Minkowski,
  Phys.\ Lett.\  B {\bf 67} (1977) 421.

\bibitem{Mohapatra:1979ia}
  R.~N.~Mohapatra and G.~Senjanovic,
  Phys.\ Rev.\ Lett.\  {\bf 44} (1980) 912.

\bibitem{Dienes:1998sb}
  K.~R.~Dienes, E.~Dudas and T.~Gherghetta,
  Nucl.\ Phys.\ B {\bf 557} (1999) 25
  [arXiv:hep-ph/9811428].


\bibitem{ArkaniHamed:1998vp}
  N.~Arkani-Hamed, S.~Dimopoulos, G.~R.~Dvali and J.~March-Russell,
  Phys.\ Rev.\  D {\bf 65} (2002) 024032
  [arXiv:hep-ph/9811448].

\bibitem{Dvali:1999cn}
  G.~R.~Dvali and A.~Y.~Smirnov,
  Nucl.\ Phys.\  B {\bf 563} (1999) 63
  [arXiv:hep-ph/9904211].

\bibitem{Faraggi:1999bm}
  A.~E.~Faraggi and M.~Pospelov,
  Phys.\ Lett.\  B {\bf 458} (1999) 237
  [arXiv:hep-ph/9901299].

\bibitem{Das:1999dx}
  A.~K.~Das and O.~C.~W.~Kong,
  Phys.\ Lett.\  B {\bf 470} (1999) 149
  [arXiv:hep-ph/9907272].

\bibitem{Mohapatra:1999zd}
  R.~N.~Mohapatra, S.~Nandi and A.~Perez-Lorenzana,
  Phys.\ Lett.\  B {\bf 466} (1999) 115
  [arXiv:hep-ph/9907520];
 R.~N.~Mohapatra and A.~Perez-Lorenzana,
  Nucl.\ Phys.\  B {\bf 576} (2000) 466
  [arXiv:hep-ph/9910474];
  R.~N.~Mohapatra and A.~Perez-Lorenzana,
  Nucl.\ Phys.\  B {\bf 593} (2001) 451
  [arXiv:hep-ph/0006278];
  R.~N.~Mohapatra, A.~Perez-Lorenzana and C.~A.~de S Pires,
  Phys.\ Lett.\  B {\bf 491} (2000) 143
  [arXiv:hep-ph/0008158];
  R.~N.~Mohapatra and A.~Perez-Lorenzana,
  Phys.\ Rev.\  D {\bf 67} (2003) 075015
  [arXiv:hep-ph/0212254];
D.~O.~Caldwell, R.~N.~Mohapatra and S.~J.~Yellin,
  Phys.\ Rev.\ Lett.\  {\bf 87} (2001) 041601
  [arXiv:hep-ph/0010353];
D.~O.~Caldwell, R.~N.~Mohapatra and S.~J.~Yellin,
  Phys.\ Rev.\  D {\bf 64} (2001) 073001
  [arXiv:hep-ph/0102279].

\bibitem{Ioannisian:1999sw}
  A.~Ioannisian and J.~W.~F.~Valle,
  Phys.\ Rev.\  D {\bf 63} (2001) 073002
  [arXiv:hep-ph/9911349].

\bibitem{Grossman:1999ra}
  Y.~Grossman and M.~Neubert,
  Phys.\ Lett.\  B {\bf 474} (2000) 361
  [arXiv:hep-ph/9912408].


\bibitem{Barbieri:2000mg}
  R.~Barbieri, P.~Creminelli and A.~Strumia,
  Nucl.\ Phys.\  B {\bf 585} (2000) 28
  [arXiv:hep-ph/0002199].

\bibitem{Ma:2000wpa}
 E.~Ma, M.~Raidal and U.~Sarkar,
  Phys.\ Rev.\ Lett.\  {\bf 85} (2000) 3769
  [arXiv:hep-ph/0006046];
E.~Ma, G.~Rajasekaran and U.~Sarkar,
  Phys.\ Lett.\  B {\bf 495} (2000) 363
  [arXiv:hep-ph/0006340].

\bibitem{McLaughlin:2000zf}
  G.~C.~McLaughlin and J.~N.~Ng,
  Phys.\ Lett.\  B {\bf 493} (2000) 88
  [arXiv:hep-ph/0008209];
C.~S.~Lam and J.~N.~Ng,
  Phys.\ Rev.\  D {\bf 64} (2001) 113006
  [arXiv:hep-ph/0104129];
 C.~S.~Lam,
  Phys.\ Rev.\  D {\bf 65} (2002) 053009
  [arXiv:hep-ph/0110142].


\bibitem{Cosme:2000ib}
  N.~Cosme, J.~M.~Frere, Y.~Gouverneur, F.~S.~Ling, 
D.~Monderen and V.~Van Elewyck,
  Phys.\ Rev.\  D {\bf 63} (2001) 113018
  [arXiv:hep-ph/0010192];
J.~M.~Frere, M.~V.~Libanov and S.~V.~Troitsky,
  Phys.\ Lett.\  B {\bf 512} (2001) 169
  [arXiv:hep-ph/0012306];
J.~M.~Frere, M.~V.~Libanov and S.~V.~Troitsky,
  JHEP {\bf 0111} (2001) 025
  [arXiv:hep-ph/0110045].

\bibitem{Agashe:2000rw}
  K.~Agashe and G.~H.~Wu,
  Phys.\ Lett.\  B {\bf 498} (2001) 230
  [arXiv:hep-ph/0010117].

\bibitem{DeGouvea:2001mz}
  A.~De Gouvea, G.~F.~Giudice, A.~Strumia and K.~Tobe,
  Nucl.\ Phys.\  B {\bf 623} (2002) 395
  [arXiv:hep-ph/0107156].


\bibitem{Davoudiasl:2002fq}
  H.~Davoudiasl, P.~Langacker and M.~Perelstein,
  Phys.\ Rev.\  D {\bf 65} (2002) 105015
  [arXiv:hep-ph/0201128].


\bibitem{Hebecker:2002re}
  A.~Hebecker and J.~March-Russell,
  Phys.\ Lett.\  B {\bf 541} (2002) 338
  [arXiv:hep-ph/0205143].

\bibitem{Lukas:2000wn}
  A.~Lukas, P.~Ramond, A.~Romanino and G.~G.~Ross,
  Phys.\ Lett.\  B {\bf 495} (2000) 136
  [arXiv:hep-ph/0008049];
 A.~Lukas, P.~Ramond, A.~Romanino and G.~G.~Ross,
  JHEP {\bf 0104} (2001) 010
  [arXiv:hep-ph/0011295];
 A.~Lukas, P.~Ramond, A.~Romanino and G.~G.~Ross,
  Int.\ J.\ Mod.\ Phys.\  A {\bf 16S1C} (2001) 934.


\bibitem{Dienes:2006bu}
  K.~R.~Dienes and I.~Sarcevic,
  Phys.\ Lett.\  B {\bf 500} (2001) 133
  [arXiv:hep-ph/0008144];
K.~R.~Dienes and S.~Hossenfelder,
  Phys.\ Rev.\  D {\bf 74} (2006) 065013
  [arXiv:hep-ph/0607112].

\bibitem{Hewett:2004py}
  J.~L.~Hewett, P.~Roy and S.~Roy,
  Phys.\ Rev.\  D {\bf 70} (2004) 051903
  [arXiv:hep-ph/0404174].

\bibitem{Eisele:2006va}
  M.~T.~Eisele and N.~Haba,
  Phys.\ Rev.\  D {\bf 74} (2006) 073007
  [arXiv:hep-ph/0603158].



\bibitem{Gherghetta:2007au}
  T.~Gherghetta, K.~Kadota and M.~Yamaguchi,
  Phys.\ Rev.\  D {\bf 76} (2007) 023516
  [arXiv:0705.1749 [hep-ph]].


\bibitem{Antoniadis:1990ew}
  I.~Antoniadis,
  Phys.\ Lett.\  B {\bf 246} (1990) 377.



\bibitem{Antoniadis:1994yi}
  I.~Antoniadis, K.~Benakli and M.~Quiros,
  Phys.\ Lett.\  B {\bf 331} (1994) 313
  [arXiv:hep-ph/9403290];
 I.~Antoniadis, K.~Benakli and M.~Quiros,
  Phys.\ Lett.\  B {\bf 460} (1999) 176
  [arXiv:hep-ph/9905311];
 E.~Accomando, I.~Antoniadis and K.~Benakli,
  Nucl.\ Phys.\  B {\bf 579} (2000) 3
  [arXiv:hep-ph/9912287].

\bibitem{Scherk:1978ta}
  J.~Scherk and J.~H.~Schwarz,
  Phys.\ Lett.\  B {\bf 82} (1979) 60;
 J.~Scherk and J.~H.~Schwarz,
  Nucl.\ Phys.\  B {\bf 153} (1979) 61.


\bibitem{Pomarol:1998sd}
  A.~Pomarol and M.~Quiros,
  Phys.\ Lett.\  B {\bf 438} (1998) 255
  [arXiv:hep-ph/9806263];
  I.~Antoniadis, S.~Dimopoulos, A.~Pomarol and M.~Quiros,
  Nucl.\ Phys.\  B {\bf 544} (1999) 503
  [arXiv:hep-ph/9810410];
  A.~Delgado, A.~Pomarol and M.~Quiros,
  Phys.\ Rev.\  D {\bf 60} (1999) 095008
  [arXiv:hep-ph/9812489].

\bibitem{Barbieri:2003kn}
 R.~Barbieri, L.~J.~Hall, G.~Marandella, Y.~Nomura, T.~Okui, 
S.~J.~Oliver and M.~Papucci,
  Nucl.\ Phys.\  B {\bf 663} (2003) 141
  [arXiv:hep-ph/0208153];
  R.~Barbieri, G.~Marandella and M.~Papucci,
  Nucl.\ Phys.\  B {\bf 668} (2003) 273
  [arXiv:hep-ph/0305044].

\bibitem{Diego:2005mu}
 D.~Diego, G.~von Gersdorff and M.~Quiros,
  JHEP {\bf 0511} (2005) 008
  [arXiv:hep-ph/0505244].

%
\bibitem{DGQ}
 D.~Diego, G.~von Gersdorff and M.~Quiros,
  Phys.\ Rev.\  D {\bf 74} (2006) 055004
  [arXiv:hep-ph/0605024].

%
\bibitem{Uranga:2005wn}
  A.~M.~Uranga,
  Class.\ Quant.\ Grav.\  {\bf 22} (2005) S41.
  %

\bibitem{Eidelman:2004wy}
 W.~M.~Yao {\it et al.}  [Particle Data Group],
  J.\ Phys.\ G {\bf 33} (2006) 1.


\bibitem{GonzalezGarcia:2002dz}
  M.~C.~Gonzalez-Garcia and Y.~Nir,
  Rev.\ Mod.\ Phys.\  {\bf 75} (2003) 345
  [arXiv:hep-ph/0202058].

\end{thebibliography}
\end{document}